\documentclass{emulateapj}

\usepackage{apjfonts}
\usepackage{float}
\usepackage{ifthen}
\usepackage{natbib}
\usepackage{rotating}
\submitted{Version accepted by the Astrophysical Journal}

\def\gs{\mathrel{\raise0.35ex\hbox{$\scriptstyle >$}\kern-0.6em
\lower0.40ex\hbox{{$\scriptstyle \sim$}}}}
\def\ls{\mathrel{\raise0.35ex\hbox{$\scriptstyle <$}\kern-0.6em
\lower0.40ex\hbox{{$\scriptstyle \sim$}}}}

\newfloat{inline}{tb}{}

\makeatother \lefthead{{\sc J. E. Geach et al}} \righthead{\sc LABs
  are powered by heating, not cooling}

\begin{document}

\title{The {\it Chandra} Deep Protocluster Survey:\\Ly$\alpha$ Blobs
  are powered by heating, not cooling}

\author{J.~E.~Geach\altaffilmark{1}, D.~M.~Alexander\altaffilmark{1},
  B.~D.~Lehmer\altaffilmark{1}, Ian~Smail\altaffilmark{2},\\
  Y.~Matsuda\altaffilmark{1}, S.~C.~Chapman\altaffilmark{3,4},
  C.~A.~Scharf\altaffilmark{5}, R.~J.~Ivison\altaffilmark{6,7}, M.~Volonteri\altaffilmark{8},\\
  T.~Yamada\altaffilmark{9}, A.\ W.\ Blain\altaffilmark{10}, R.\ G.\
  Bower\altaffilmark{2}, F.\ E.\ Bauer\altaffilmark{5} \&
  A. Basu-Zych\altaffilmark{5}}

\altaffiltext{1}{Department of Physics, Durham University, South Road,
  Durham DH1 3LE, UK.  E-mail: j.e.geach@durham.ac.uk}
\altaffiltext{2}{Institute for Computational Cosmology, Department of
  Physics, Durham University, South Road, Durham DH1 3LE, UK.}
\altaffiltext{3}{Institute of Astronomy, Madingley Road, Cambridge CB3
  0HA, UK} \altaffiltext{4}{Department of Physics and Astronomy,
  University of Victoria, Victoria, B.C., V8P 1A1, Canada}
\altaffiltext{5}{Columbia Astrophysics Laboratory, Columbia
  University, Pupin Laboratories, 550 W. 120th St., Rm 1418, New York,
  NY 10027, USA} \altaffiltext{6}{SUPA, Institute for Astronomy, Royal
  Observatory of Edinburgh, Blackford Hill, Edinburgh, EH9 3HJ, UK}
\altaffiltext{7}{Astronomy Technology Centre, Royal Observatory of
  Edinburgh, Blackford Hill, Edinburgh, EH9 3HJ, UK}
\altaffiltext{8}{Department of Astronomy, University of Michigan, Ann
  Arbor, MI, USA} \altaffiltext{9}{National Astronomical Observatory
  of Japan, Tokyo 181-8588, Japan} \altaffiltext{10}{Department of
  Astronomy, California Institute of Technology, MC 105-24, 1200, East
  California Blvd, Pasadena, CA, 91125. USA. }

\label{firstpage}

\begin{abstract}
  We present the results of a 400\,ks {\it Chandra} survey of 29
  extended Ly$\alpha$ emitting nebulae (Ly$\alpha$ Blobs, LABs) in the
  $z=3.09$ proto-cluster in the SS\,A22 field. We detect luminous
  X-ray counterparts in five LABs, implying a large fraction of active
  galactic nuclei (AGN) in LABs, $f_{\rm AGN} = 17^{+12}_{-7}$\% down
  to $L_{\rm 2-32keV} \sim 10^{44}$\,erg\,s$^{-1}$. All of the AGN
  appear to be heavily obscured, with spectral indices implying
  obscuring column densities of $N_{\rm H} > 10^{23}$\,cm$^{-2}$. The
  AGN fraction should be considered a lower limit, since several more
  LABs not detected with {\it Chandra} show AGN signatures in their
  mid-infrared emission. We show that the UV luminosities of the AGN
  are easily capable of powering the extended Ly$\alpha$ emission via
  photo-ionization alone. When combined with the UV flux from a
  starburst component, and energy deposited by mechanical feedback, we
  demonstrate that `heating' by a central source, rather than
  gravitational cooling is the most likely power source of LABs. We
  argue that all LABs could be powered in this manner, but that the
  luminous host galaxies are often just below the sensitivity limits
  of current instrumentation, or are heavily obscured. No individual
  LABs show evidence for extended X-ray emission, and a stack
  equivalent to a $\gs$9\,Ms exposure of an average LAB also yields no
  statistical detection of a diffuse X-ray component. The resulting
  diffuse X-ray/Ly$\alpha$ luminosity limit implies there is no hot
  ($T\gs10^7$\,K) gas component in these halos, and also rules out
  inverse Compton scattering of cosmic microwave background photons,
  or local far-infrared photons, as a viable power source for LABs.
\end{abstract}


\begin{table*}
  \caption{X-ray properties of LABs in SSA\,22.}
\begin{tabular*}{0.99\textwidth}{@{\extracolsep{\fill}}lccccccccl}
  \hline
  LAB ID  & $\alpha_{\rm J2000}$ & $\delta_{\rm J2000}$ & $f_{\rm
    0.5-2\,keV}$ & $f_{\rm 2-8\,keV}$ & $f_{\rm 0.5-8\,keV}$ & $L_{\rm 2-32\,keV}$ & $\Gamma_{\rm eff}$ &Offset & Note\cr
  & (h m s) & ($\circ$ $'$ $''$) &
  \multicolumn{3}{c}{(10$^{-16}$\,erg\,s$^{-1}$\,cm$^{-2}$)} &
  ($10^{44}$\,erg\,s$^{-1}$) && ($''$)\cr
  \hline
 \cr
   \multicolumn{10}{c}{X-ray detected LABs}\cr
\cr
  LAB2 &  22 17 39.00 &  +00 13 27.5 &  $<$1.45 & 12.40$\pm$0.48 &
  9.64$\pm$0.34 & 0.81$\pm$0.03& $<$0.42 & 3.46$\pm$0.80& SMG /
  8$\mu$m detected\cr
  LAB3 &  22 17 59.10 &  +00 15 28.0 & 6.91$\pm$0.11 & 18.00$\pm$0.40& 25.50$\pm$0.22 & 2.13$\pm$0.02 & 1.28$^{+0.30}_{-0
  .28}$ & 2.59$\pm$0.37\cr
LAB12 &         22 17 31.90 &  +00 16 58.0 &
0.80$\pm$0.11&10.30$\pm$0.50 & 10.90$\pm$0.37 & 0.91$\pm$0.03
&0.17$^{+0.52}_{-0.53}$  & 2.86$\pm$0.80 & 24$\mu$m detected\cr
LAB14 &         22 17 35.90 &  +00 15 58.0 & 5.07$\pm$0.09&16.50$\pm$0.37 & 21.70$\pm$0.21 & 1.82$\pm$0.02&1.13$^{+0.27
}_{-0.25}$ & 1.47$\pm$0.30 & SMG / 24$\mu$m detected\cr
LAB18 &         22 17 28.90 &  +00 07 51.0 & $<$3.28 & 20.80$\pm$0.46 & 19.00$\pm$0.32 & 1.59$\pm$0.03 &$<$0.63  &7.02$
\pm$1.41 & SMG / 24$\mu$m detected\cr
 \cr
   \multicolumn{10}{c}{X-ray non-detected LABs}\cr
\cr
LAB1 & 	22 17 26.00 & 	+00 12 36.6 & $<$2.90 & $<$1.99 & $<$5.10 &
$<$0.24 & -- & -- & SMG / 8$\mu$m detected\cr
LAB4 & 	22 17 25.10 & 	+00 22 10.0 & $<$6.76 & $<$3.71 & $<$10.44 & $<$0.56 & -- & -- &  \cr
LAB5 & 	22 17 11.70 & 	+00 16 43.3 & $<$5.31 & $<$2.75 & $<$8.87 &
$<$0.44 & -- & -- &  SMG / 8$\mu$m detected\cr
LAB7 & 	22 17 41.00 & 	+00 11 26.0 & $<$2.68 & $<$1.90 & $<$4.28 & $<$0.22 & -- & -- &  \cr
LAB8 & 	22 17 26.10 & 	+00 12 53.0 & $<$2.35 & $<$1.37 & $<$4.87 & $<$0.20 & -- & -- &  \cr
LAB9 & 	22 17 51.00 & 	+00 17 26.0 & $<$4.49 & $<$2.80 & $<$7.07 & $<$0.37 & -- & -- &  \cr
LAB11 & 	22 17 20.30 & 	+00 17 32.0 & $<$3.42 & $<$1.59 & $<$6.22 & $<$0.28 & -- & -- &  \cr
LAB13 & 	22 18 07.90 & 	+00 16 46.0 & $<$18.92 & $<$10.12 & $<$31.04 & $<$1.57 & -- & -- &  \cr
LAB15 & 	22 18 08.30 & 	+00 10 21.0 & $<$10.46 & $<$4.31 & $<$16.92 & $<$0.87 & -- & -- &  \cr
LAB16 & 	22 17 24.80 & 	+00 11 16.0 & $<$4.34 & $<$2.68 &
$<$6.49 & $<$0.36 & -- & -- &  24$\mu$m / 8$\mu$m detected\cr
LAB19 & 	22 17 19.50 & 	+00 18 46.0 & $<$4.38 & $<$1.76 & $<$8.33 & $<$0.36 & -- & -- &  \cr
LAB20 & 	22 17 35.30 & 	+00 12 48.0 & $<$2.67 & $<$1.96 & $<$4.13 & $<$0.22 & -- & -- &  \cr
LAB21 & 	22 18 17.30 & 	+00 12 08.0 & $<$34.26 & $<$18.60 & $<$54.54 & $<$2.84 & -- & -- &  \cr
LAB22 & 	22 17 34.90 & 	+00 23 35.0 & $<$6.72 & $<$3.45 & $<$10.82 & $<$0.56 & -- & -- &  \cr
LAB24 & 	22 18 00.90 & 	+00 14 40.0 & $<$4.71 & $<$2.14 & $<$8.67 & $<$0.39 & -- & -- &  \cr
LAB25 & 	22 17 22.50 & 	+00 15 50.0 & $<$2.72 & $<$2.00 & $<$4.85 & $<$0.23 & -- & -- &  \cr
LAB26 & 	22 17 50.40 & 	+00 17 33.0 & $<$2.81 & $<$1.32 & $<$5.37 & $<$0.23 & -- & -- &  \cr
LAB27 & 	22 17 06.90 & 	+00 21 30.0 & $<$11.68 & $<$6.33 & $<$17.45 & $<$0.97 & -- & -- &  \cr
LAB28 & 	22 17 59.20 & 	+00 22 53.0 & $<$11.88 & $<$5.26 & $<$18.94 & $<$0.99 & -- & -- &  \cr
LAB30 & 	22 17 32.40 & 	+00 11 33.0 & $<$3.25 & $<$2.32 & $<$5.31 & $<$0.27 & -- & -- &  \cr
LAB31 & 	22 17 38.90 & 	+00 11 01.0 & $<$2.80 & $<$1.65 & $<$5.17 & $<$0.23 & -- & -- &  \cr
LAB32 & 	22 17 23.80 & 	+00 21 55.0 & $<$5.64 & $<$3.07 & $<$9.06 & $<$0.47 & -- & -- &  \cr
LAB33 & 	22 18 12.50 & 	+00 14 32.0 & $<$25.07 & $<$11.88 & $<$42.09 & $<$2.08 & -- & -- &  \cr
LAB35 & 	22 17 24.80 & 	+00 17 17.0 & $<$3.27 & $<$1.83 & $<$5.91 & $<$0.27 & -- & -- &  \cr
  \hline
\end{tabular*}
\begin{minipage}{\textwidth}\vspace{0.3cm}
  Notes --- Co-ordinates correspond to the centroid of X-ray
  detection. X-ray fluxes are in the observed frame, but the full band
  luminosity is quoted in the 2--32\,keV rest-frame; X-ray properties
  are from Lehmer et al.\ (2009). $\Gamma_{\rm eff}$ is the inferred
  effective photon index. `Offset' refers to the angular separation
  between X-ray centroid and peak of Ly$\alpha$ emission (errors
  reflects 1$\sigma$ uncertainty in X-ray position).  \end{minipage}
\end{table*}

\section{Introduction}

It appears that feedback between galaxies and the intergalactic medium
(IGM) plays a significant role in the formation and evolution of
galaxies (Bower et al.\ 2006; Croton et al.\ 2006). Without it, even
some of the basic properties of galaxies (such as stellar mass) cannot
be re-produced in current models of galaxy formation. Gas cooling
within dark matter halos is countered by outflows from starbursts and
active galactic nuclei (AGN) and other heating mechanisms. These not
only heat, but can also enrich the intergalactic medium (IGM), and
truncate star formation within the host galaxies -- preventing a glut
of $>$$L_\star$ galaxies in the local Universe. Placing empirical
constraints on these processes, and understanding their detailed
physics, is therefore of vital importance.

Recently there has been great interest in the highly extended
($\sim$30--200\,kpc in projected linear extent) Ly$\alpha$
line-emitting nebulae ($L_{\rm
  Ly\alpha}\sim10^{43-44}$\,erg\,s$^{-1}$) identified in high-redshift
narrowband surveys: `Ly$\alpha$ Blobs' (LABs) (Fynbo et al.\ 1999;
Keel et al.\ 1999; Steidel et al.\ 2000; Francis et al.\ 2001; Palunas
et al.\ 2004; Matsuda et al.\ 2004; Dey et al.\ 2005; Smith et al.\
2008).  The most important questions in LAB studies remain unanswered:
how are they formed and what maintains their power? One of the main
reasons that these objects have aroused curiosity is the possibility
that they trace feedback events during the formation of massive
galaxies (Chapman et al.\ 2001; Geach et al.\ 2005, 2007; Webb et al.\
2009), but we still lack a definitive model of LAB formation.

What are the possible formation mechanisms of LABs? At first glance,
these objects appear to be good candidates for the Ly$\alpha$ `fuzz'
predicted to exist around primordial galaxies in simple models of
galaxy formation (e.g.\ Rees \& Ostriker\ 1977; Haiman\ et al.\ 2000;
Haiman \& Rees\ 2001; Birnboim \& Dekel\ 2003). Cooling of pristine
gas within a dark matter halo via Ly$\alpha$ emission could, in part,
provide the energy required to power a LAB via the release of
gravitational potential energy (e.g.\ Fardal et al.\ 2001; Nilsson et
al.\ 2006; Smith \& Jarvis\ 2007).  However, this has to be reconciled
with the fact that many LABs appear to be associated with extremely
luminous galaxies (Chapman et al.\ 2001; Dey et al.\ 2005; Geach et
al.\ 2005, 2007; Colbert et al.\ 2006; Beleen et al.\ 2008; Webb et
al.\ 2009) with bolometric luminosities several orders of magnitude
greater than that of the Ly$\alpha$ emission. Therefore, some models
of LAB formation propose a `heating' scenario, where the energy
release associated with intense star formation or AGN within the LABs'
host galaxies powers the extended line emission (e.g.\ Ohyama et al.\
2003). It has also been postulated that inverse Compton scattered
cosmic microwave background (CMB) photons could go on to photo-ionize
a neutral gas halo (e.g.\ Fabian et al.\ 2009). This mechanism is
thought to give rise to extended X-ray emission around luminous radio
galaxies at $z>2$ (Scharf et al.\ 2003). Unfortunately the current
limits on the soft, diffuse X-ray emission around LABs are poor.

In this paper we concentrate on identifying the power sources of 29
LABs in the SSA\,22 proto-cluster (Steidel et al.\ 2000; Hayashino et
al.\ 2004): a region $\sim$6$\times$ over-dense compared to the field
at $z=3.09$, and containing the richest association of LABs known
(Matsuda et al.\ 2004). Our aim is to identify both un-obscured and
obscured AGN within LABs, and also search for evidence of extended
X-ray emission which could imply inverse Compton scattering, or a hot
(few keV) gas component in the extended halos. Understanding the
importance of AGN in LABs' host galaxies is crucial to assess whether
the feedback physics associated with black-hole growth is powering the
extended Ly$\alpha$ emission. To do this we exploit a very deep
($\sim$400\,ks) X-ray exposure: the {\it Chandra} Deep Protocluster
Survey (Lehmer et al.\ 2009).

Throughout this work we assume a cosmology where $(\Omega_{\rm m},
\Omega_\Lambda) = (0.3,0.7)$ and $H_0
=70$\,km\,s$^{-1}$\,Mpc$^{-1}$. At $z=3.09$ this corresponds to a
luminosity distance of 26.3\,Gpc and scale of
7.6\,kpc/$''$. Magnitudes are all on the AB scale, and all X-ray
fluxes have been corrected for Galactic absorption; the Galactic H{\sc
  i} column density towards SSA\,22 is $N_H =
4.6\times10^{20}$\,cm$^{-2}$ (Stark et al.\ 1992).

%FIGURE 1

\begin{figure*}
\centerline{\includegraphics[width=\textwidth]{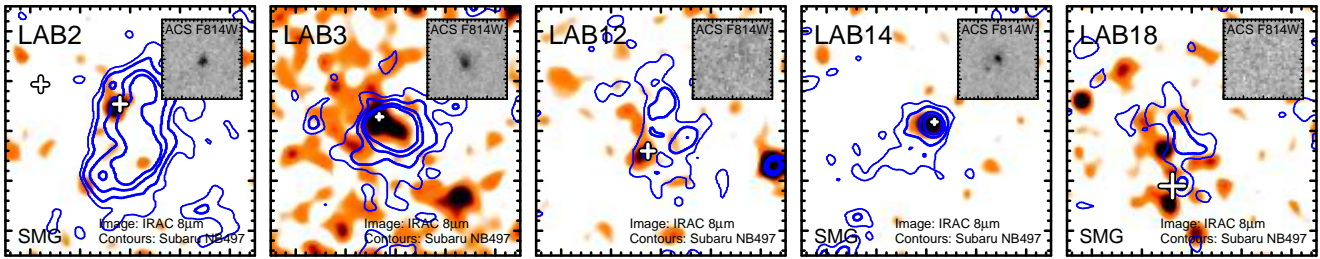}}
\vspace{1mm}
\caption{Thumbnail images of the X-ray detected LABs. The main panels
  show the full extent of the LABs ($30''\times30''$\,arcsec or $230
  \times 230$\,kpc). The background image shows the {\it SST} IRAC
  8$\mu$m emission, and we indicate the position of the X-ray
  counterparts as crosses. The sizes of the crosses correspond to the
  1$\sigma$ uncertainties in the X-ray positions. The contours
  represent Ly$\alpha$ emission traced by the Subaru NB497 (continuum
  corrected) narrowband imaging of Matsuda et al.\ (2004) and are
  spaced at (5,\ 10,\ 20,\
  30)$\times10^{-20}$\,erg\,s$^{-1}$\,cm$^{-2}$\,arcsec$^{-2}$.  The
  inset images show {\it HST} ACS F814W (rest-frame UV) postage stamps
  extracted at the location of the X-ray source ($2''\times2''$,
  $15\times 15$\,kpc).  It is interesting to note that none of the
  LABs in this sample are symmetric about the X-ray emission -- often
  the Ly$\alpha$ emission is extended away from the active source.}
\end{figure*}

\section{Observations}

A 330\,arcmin$^2$ region in the SSA\,22 field was observed for
$\sim$400\,ks using the ACIS camera on-board {\it Chandra} (P.\ I.:
D.\ M.\ Alexander). The observations comprise of four {\it Chandra}
pointings taken between 2007 October 1 and 2007 December 30
(Obs. I.D.s 8034, 8035, 8036, 9717), centred on the LBG survey region
of Steidel et al.\ (2003), 22\ 17\ 36, +00\ 15\ 33 (J2000.0). These
observations cover 29 of the 35 SSA\,22 LABs of Matsuda et al.\
(2004); only LAB\,6,\ 10,\ 17,\ 21,\ 23,\ 29 are {\it not} covered by
the {\it Chandra} observations.

Slight differences in roll angle between the four observations results
in a total survey area $\sim$12\% larger than a single ACIS-I field of
view ($16.9'\times16.9'$), and the variation in effective exposure
time across the map is taken into account in the subsequent source
extraction.  A full description of data reduction, source detection
and catalogue creation can be found in Lehmer et al.\ (2009). In
summary the survey reaches a point-source sensitivity limit of
$4.8\times10^{-17}$\,erg\,s$^{-1}$\,cm$^{-2}$ and
$2.7\times10^{-16}$\,erg\,s$^{-1}$\,cm$^{-2}$ in the 0.5--2\,keV and
2--8\,keV bands respectively. At the redshift of the protocluster,
these correspond to luminosities of $3.7\times10^{42}$\,erg\,s$^{-1}$
and $2.1\times10^{43}$\,erg\,s$^{-1}$ at rest-frame energies of
2--8\,keV and 8--32\,keV respectively.

The SSA\,22 region was surveyed by {\it HST} Advanced Camera for
Surveys (ACS) in a sparse mosaic of 10 pointings during August 2005 (3
orbits per pointing, $\sim6.2$ks. P.\ I.:\ S.\ C.\ Chapman, P.I.D.\
10405). A single filter, F814W, was used -- probing rest-frame
$\sim$2000\AA\ emission at the redshift of the protocluster. Data was
reduced using the standard Space Telescope Science Institute software
{\sc Multidrizzle}.  We have also obtained an additional three ACS
pointings from the Gemini Deep Deep Survey (GDDS) archive -- again
this was reduced from the archive `flat' stage using {\sc
  Multidrizzle}. Since the {\it HST} mosaic is sparse, 14 LABs in the
{\it Chandra} map do not have ACS coverage, but since this is not a
comprehensive morphological study, this does not impact our analysis
of the AGN properties of LABs.

The SSA\,22 field has been imaged with {\it Spitzer Space Telescope}
IRAC 3.6, 4.5, 5.8 and 8.0$\mu$m imaging as part of GO program \#64
and GTO program \#30328.  The data has been described and presented in
Webb et al.\ (2009). In summary, there is uniform coverage of 225\
arcmin$^2$ in all four IRAC bands with an integration time of 7.5\
ks/pix. Unless otherwise stated, the IRAC photometry presented in this
work has been taken from Webb et al.\ (2009), with fluxes measured in
3.4$''$ diameter apertures, corrected to total fluxes. The region
covered by IRAC imaging also has MIPS 24$\mu$m coverage (from the same
{\it SST} programs), with an integration time of 1.2\,ks/pix. The MIPS
data is also discussed in Webb et al.\ (2009). Of all the {\it
  Chandra} covered LABs, only LAB\,28 is not covered by the
mid-infrared imaging.

In this work we also make use of archival UKIDSS-Deep eXtragalactic
Survey (DXS)\footnote{{\tt
    http://www.ukidss.org/surveys/surveys.html}} {\it J/K}-band
imaging of SSA\,22. In addition to the DXS imaging, we have
supplemented the near-IR coverage with UKIRT/WFCAM {\it H}-band
imaging of the {\it Chandra} field. This data was obtained in
UKIRT/WFCAM service mode (project U/SERV/1759) and reduced using our
in-house WFCAM data reduction pipeline (see Geach et al.\ 2008 for
details). The {\it H}-band imaging was taken in moderate
seeing, $\ls$1$''$, and reaches a 3$\sigma$ depth of
$\sim$21.5\,mag. For comparison, the equivalent depth of the DXS
imaging is 22.0\,mag and 21.7\,mag in {\it J}- and {\it K-}bands
respectively.

\section{Results}

\subsection{Identifying AGN in LABs}

\subsubsection{X-ray counterparts}

To identify X-ray sources associated with the LABs we first identify
all X-ray counterparts within a radius $2R_{\rm LAB}$ of the peak of
the Ly$\alpha$ emission. The effective LAB radius is defined by the
isophotal area: $R_{\rm LAB}= (A_{\rm LAB}/\pi)^{1/2}$ (we assume the
isophotal areas from Matsuda et al.\ 2004). We find unambiguous X-ray
counterparts to five LABs: LAB\,2 (previously identified in a 78\,ks
{\it Chandra} exposure by Basu-Zych \& Scharf\ 2004), LAB\,3, 12, 14,
\& 18; see Table~1. In Figure~1 we present thumbnail images of the
X-ray detected LABs, indicating the position of the X-ray detection
relative to the Ly$\alpha$ emission. As can be seen, often the X-ray
counterpart is slightly offset from the peak of the Ly$\alpha$
emission.

All five of the X-ray detected LABs are covered by the {\it HST}/ACS
mosaic. LAB\,2. LAB\,3 and LAB\,14 all have compact rest-frame UV
morphologies, although LAB\,14 has some evidence of a
merger/interaction, with two components separated on a scale of
$\lesssim$0.5$''$. Interestingly, the alignment of these two
components is in the same direction as the extended Ly$\alpha$
emission. LAB\,12 and LAB\,18 have no counterpart in the ACS image,
and this could reflect more extended, low surface brightness continuum
emission in these LABs (c.f.\ LAB\,1, Chapman et al.\ 2004). We
discuss the multi-wavelength properties of the LABs further in \S3.2.

In order to estimate the contamination rate from chance alignments of
X-ray detections with LABs, we calculate the probability of finding a
$L_X >L_{X, \rm LAB}$ association by randomly placing an aperture of
radius $R_{\rm LAB}$ on the X-ray map and counting the number of
`detections' within it. We repeat this process 1000 times for each LAB
to build-up a statistical representation of the robustness of each
detection. The resulting probability of randomly associating an X-ray
counterpart with a LAB is 10\%, and so we expect 0.5 false
matches. This contamination factor is dominated by the three largest
LABs in the survey. For example, if one excludes them, this
contamination drops by a factor 2.  Assuming the X-ray detections
pin-point AGN in these five LABs, we measure the luminous AGN fraction
in LABs in SSA\,22 to be $f_{\rm AGN} = 17^{+12}_{-7}$\% (Gehrels\
1986). This fraction should be considered a lower limit because we
have only considered X-ray luminous AGN. In the following section we
examine the potential for detecting obscured AGN within the remaining
LABs.

% FIGURE 2

\begin{figure}
\centerline{\includegraphics[width=0.5\textwidth]{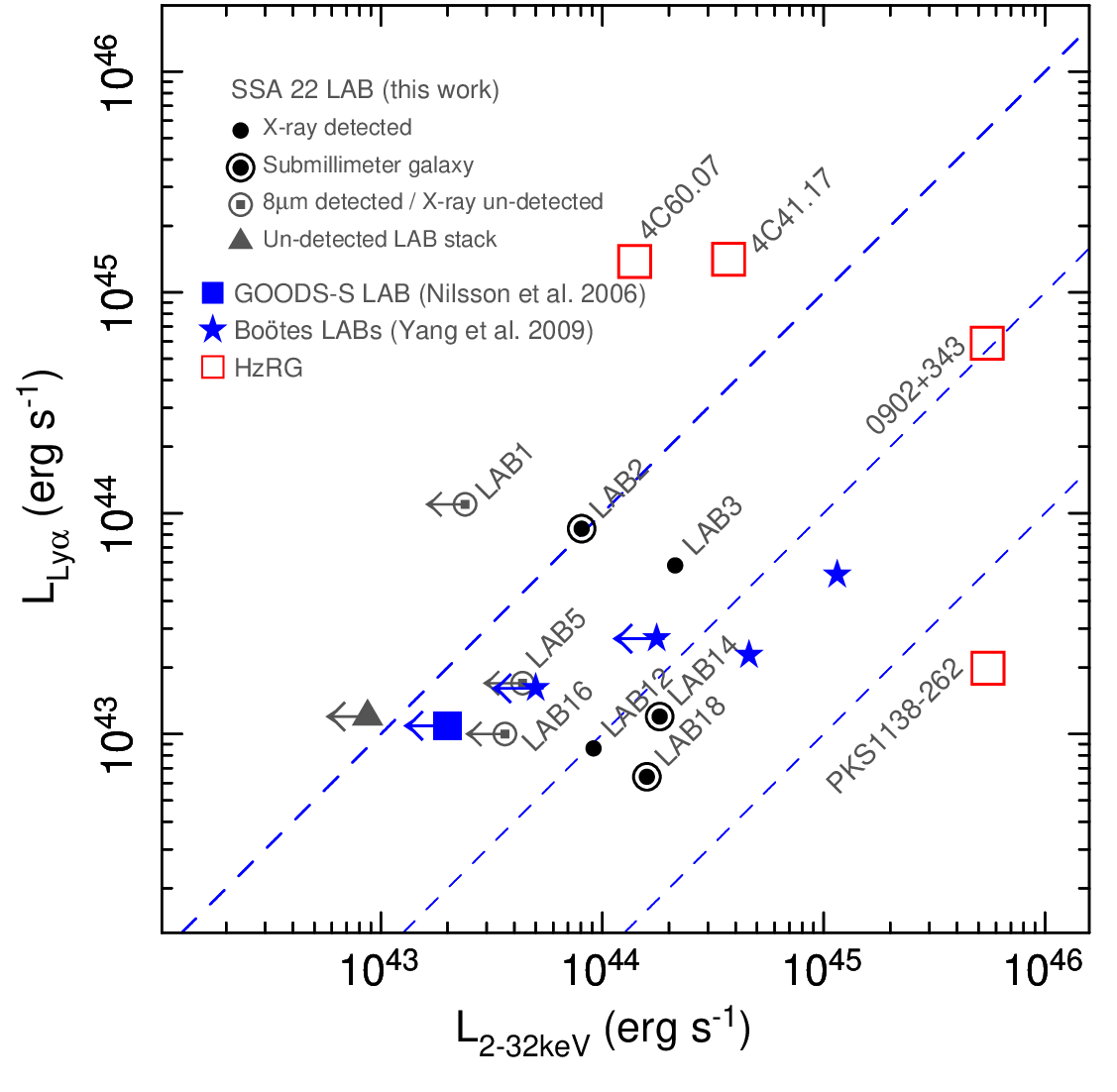}}
\caption{A comparison of Ly$\alpha$ and X-ray (observed 0.5--8\,keV,
  rest-frame 2--32\,keV) luminosity for LABs and HzRGs. For LABs not
  formally detected in the X-ray image, we indicate 3$\sigma$ upper
  limit for a stack of 21 LABs (excluding the formally detected LABs
  and those containing 8$\mu$m counterparts). We also show the
  3$\sigma$ upper limits for three 8$\mu$m detected LABs with evidence
  of ULIRG-like SEDs (Geach et al.\ 2007; Webb et al.\ 2009, \S3.1.2).
  Note that the LAB detected by Nilsson et al.\ (2006) in GOODS-S has
  a $L_X/L_{\rm Ly\alpha}$ limit not inconsistent with the range
  observed in the SSA\,22 LABs. The GOODS-S LAB is proposed to be the
  best candidate for a LAB powered by cooling flows (see \S4.4 for a
  discussion).  For comparison to LABs, we show the positions of four
  HzRGs (Reuland et al.\ 2003), which also exhibit large Ly$\alpha$
  halos. The most notable difference between LABs and the nebulae
  around HzRGs is that although both populations span a similar range
  of $L_X/L_{\rm Ly\alpha}$, HzRGs are 10--100$\times$ more luminous
  in terms of both their Ly$\alpha$ {\it and} X-ray luminosity.}
\end{figure}

\subsubsection{Searching for X-ray un-detected AGN in LABs}

Enshrouding an AGN with gas and dust could render it un-detectable
even in our deep X-ray survey. Nevertheless, we can potentially
identify these systems by turning to mid-infrared observations. Dust
heated by the AGN gives rise to a steep power-law ($S_\nu \propto
\nu^{-\alpha}$) continuum in the rest-frame near-infrared, in excess
of that expected from a stellar continuum. At $z=3.09$ the IRAC
8$\mu$m imaging is probing rest-frame $\sim$2$\mu$m emission beyond
the peak of the stellar continuum at 1.6$\mu$m. It is therefore ideal
for identifying AGN (Lacy et al.\ 2004).

All five X-ray detected AGN are associated with 8$\mu$m sources
(although LAB\,3 suffers some confusion from a nearby foreground
source). Webb et al.\ do not associate LAB\,12 with an 8$\mu$m
counterpart; however, we find a fairly low significance
($\lesssim$5$\sigma$) source coincident with the X-ray point source in
LAB\,12 (Fig~1). In addition to these unambiguous AGN, Geach et al.\
(2007) identified LAB\,1 with an 8$\mu$m counterpart, and Webb et al.\
(2009) detect 8$\mu$m counterparts in two other LABs: LAB\,5 and
LAB\,16. Although LAB\,1, 5 and 16 are not detected at X-ray energies,
LAB\,1 and LAB\,5 are 850$\mu$m emitters (submillimeter galaxies
[SMGs] Chapman et al.\ 2001; 2004, Geach et al.\ 2005) and LAB\,16 is
detected at 24$\mu$m (Webb et al.\ 2009). These mid- and far-infrared
detections link these LABs to energetic, but dusty, power-sources.

Are these 8$\mu$m detections likely to be obscured AGN? Webb et al.\
(2009) show that {\it all} of the 8$\mu$m-detected LABs have
rest-frame near-infrared colours consistent with an AGN or ULIRG
SED. To examine the possibility that LAB\,1, LAB\,5 and LAB\,16 host
heavily obscured AGN (or low-luminosity AGN below the detection limit)
we stack the X-ray map at these three positions using the technique
outlined in Lehmer et al.\ (2008). We find a marginally significant
(93.6\% confidence) excess of 6.6 counts compared to 3.5 expected from
the background. This corresponds to an average X-ray luminosity of
$\left<L_{\rm 2-32\,keV}\right> \simeq
1.5\times10^{43}$\,erg\,s$^{-1}$. This is only marginally significant,
and the 3$\sigma$ upper limit for this stack is $L_{\rm 2-32\,keV} <
4.9\times10^{43}$\,erg\,s$^{-1}$.  In comparison, the stacked X-ray
counts from all remaining 21 LABs covered by the {\it Chandra}
exposure yields no significant detection, with a 3$\sigma$ upper limit
of $L_{\rm 2-32\,keV} < 9.2\times10^{42}$\,erg\,s$^{-1}$. The stacking
position for each of these LABs is taken as the position of the peak
Ly$\alpha$. Although Fig.~1 shows that the AGN does not have to be
located at the centre of the Ly$\alpha$ emission, the influence of
this offset is less important for the majority of LABs, which have
relatively small spatial extents.

We re-iterate that given the presence of a hidden population of AGN in
LABs, the AGN fraction derived in \S\S3.1.1 should be considered a
lower limit. If one includes LAB\,1, LAB\,5 and LAB\,16, the AGN
fraction could be as large as $28^{+14}_{-10}$\%. Such a large AGN
fraction hints that there is a strong link between the active host
galaxy and the presence of an extended Ly$\alpha$ halo. Our results
support the findings of Yang et al.\ (2009), who identify two bright
AGN in four of the LABs they detect in the NOAO Deep Wide Field Survey
Bo\"otes field. Comparison of AGN fractions between surveys is
complicated by the slightly different selection criteria. If we adjust
our SSA\,22 LAB sample to reflect the Yang et al. (2009) LAB selection
criteria, then we find an AGN fraction of 44$^{+35}_{-21}$\%,
consistent with the 50\% fraction in Bo\"otes.

\begin{figure*}
\centerline{\includegraphics[width=0.70\textwidth]{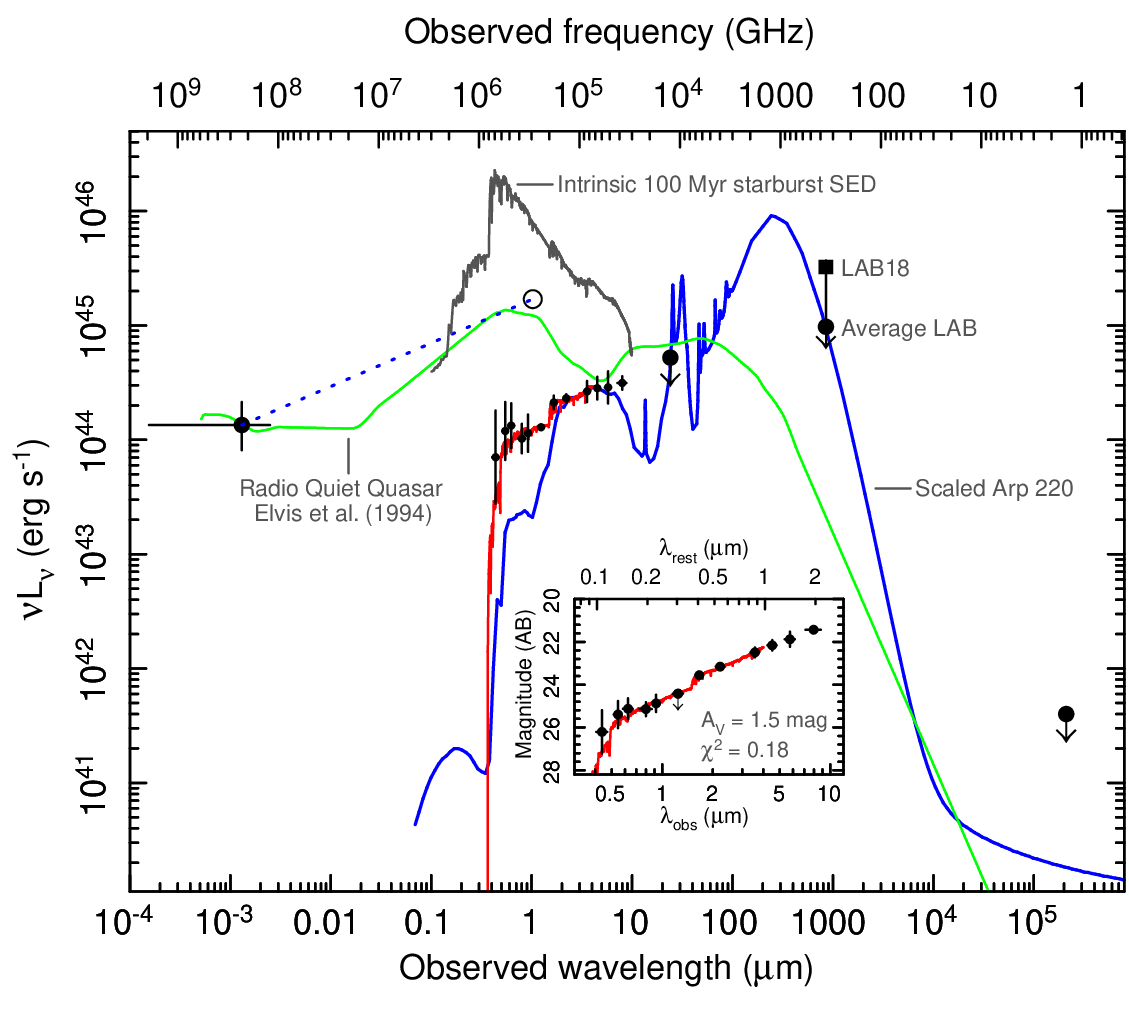}}
\caption{The composite spectral energy distribution of the X-ray
  detected LABs in SSA\,22. Where shown, upper limits are at the
  3$\sigma$ level, and where relevant we have shown the range of
  luminosities in the sample to indicate variations from
  source-to-source. As a guide, we show the SED of Arp\ 220 (Silva et
  al.\ 1998) redshifted to $z=3.09$ and normalized to our observed
  4.5$\mu$m luminosity. For comparison, we also show the radio quiet
  quasar (RQQ) template of Elvis et al.\ (1994) redshifted and scaled
  to our average X-ray flux. The UV luminosity predicted by the RQQ
  template is in good agreement with the X-ray/UV power-law
  extrapolation of Steffen et al.\ (2006) which we indicate as a
  dotted line and point at $\lambda=2500$\AA. In the inset we show a
  fit to the optical--near-infrared photometry using {\sc hyperz}. The
  fit is a moderately reddened ($A_V\sim1.5$\,mag) continuous star
  formation history of age $\sim$100\,Myr. This this is to be compared
  with the {\it intrinsic} UV luminosity from the AGN and starburst
  component (in the main panel we show the intrinsic SED of a 100\,Myr
  old starburst, normalised to the SFR estimated from the far-infrared
  emission). Note that the intrinsic UV luminosity predicted for the
  starburst and AGN components are orders of magnitude larger than the
  Ly$\alpha$ luminosity of LABs.}
\end{figure*}

\subsection{Properties of the AGN}

The properties of the five X-ray counterparts to LABs (and upper
limits for the non-detections) are summarized in Table\ 1. All five
LABs have rest-frame 2--32\,keV luminosities of $L_{\rm 2-32keV}
\sim10^{44}$\,erg\,s$^{-1}$, and hard effective photon indices
($\Gamma_{\rm eff}\ls 1$), implying intrinsic column densities of
order $N_{\rm H} \gs 10^{23}$\,cm$^{-2}$ (see Figure\ 3 of Alexander
et al.\ 2005). The average 3$\sigma$ upper limits for un-detected LABs
are $f_{\rm 0.5-2\,keV}<2.3\times10^{-16} $\,erg\,s$^{-1}$\,cm$^{-2}$
and $f_{\rm 2-8\,keV}<7.1\times10^{-16} $\,erg\,s$^{-1}$\,cm$^{-2}$.
This corresponds to a luminosity limit of $L_{\rm 2-32\,keV} <
3.6\times10^{43}$\,erg\,s$^{-1}$. Note that there is slight variation
in the limits over the field due to the varying exposure time across
the map. 

In Figure~2 we compare the X-ray luminosities of the LABs to their
Ly$\alpha$ luminosities, which show that $L_{\rm 2-32\,keV}>L_{\rm
  Ly\alpha}$ with a similar range of $L_{\rm 2-32\,keV}/L_{\rm
  Ly\alpha}$ as high-redshift radio galaxies (HzRGs, Reuland et al.\
2003). By comparison, HzRGs are generally 10--100$\times$ more
luminous in both Ly$\alpha$ luminosity and X-ray luminosity, and so it
is not clear if LABs are simply `scaled down' versions of the
Ly$\alpha$ halos around HzRGs, but it is clear that both populations
are characterised by bolometrically luminous galaxies.  This suggests
that the LAB phenomenon could be an important, and perhaps ubiquitous
phase in the formation of massive galaxies in general. Note that
compared to surveys of radio galaxies, wide-field surveys of LABs (and
more importantly, comprehensive multi-wavelength follow-up) have yet
to cover significant volumes needed to identify the most extreme
examples. Clearly, larger samples of LABs are required to provide a
wide dynamic range in properties to properly assess their relation to
other high-$z$ galaxy populations.

What are the multi-wavelength properties of these AGN LAB hosts? In
Figure~3 we present the composite SED of the X-ray detected LABs,
covering X-ray to radio wavelengths. As a guide, we compare the
observed photometry to two representative SEDs: the archetypal local
ULIRG Arp\ 220 (Silva et al.\ 1998), and the radio quiet quasar (RQQ)
template of Elvis et al.\ (1994).  LAB\,2, 14 and 18 contain SMGs
(Chapman et al.\ 2001; Geach et al.\ 2005), and we indicate their
range in luminosity (as well as an upper limit for non-detections) on
Figure~3. Note that Geach et al.\ (2005) showed that LABs not formally
detected at 850$\mu$m have a statistical signature of submillimeter
emission at the $\sim$3\,mJy level. From the sub-mm flux, we can
estimate the galaxies' far-infrared luminosities, $L_{\rm FIR}$. We
model the far-infrared emission as a $T$-$\alpha$-$\beta$ modified
black-body\footnote{Here $\alpha$ describes the power-law Wien tail in
  the mid-infrared, $\beta$ describes the emissivity in the
  Rayleigh-Jeans regime and the temperature $T$ controls the frequency
  of the peak of the spectrum. If $\alpha$ and $\beta$ are fixed, then
  at $z=3.09$ the sub-mm to far-infrared conversion varies like:
  $L_{\rm FIR}/L_{\odot} = 7.2\times10^7(T/ 1\,{\rm K})^{2.66}(S_{\rm
    850} / 1\,{\rm mJy})$ over the range $30<T<50$\,K. Parameterising
  $L_{\rm FIR}$ in this way allows the reader to re-scale our
  luminosity estimates for alternative temperatures.} (Blain et al.\
2003).  Assuming $(T,\alpha, \beta) = (35\,{\rm K},4,1.5)$, the
far-infrared luminosities of the LABs are in the ultraluminous regime,
with $L_{\rm FIR}\gs2.5\times10^{12}L_\odot$ (slightly more
conservative than presented in Geach et al.\ [2005]).

The LABs containing formally detected SMGs have implied $L_X/L_{\rm
  FIR}\sim0.003$--$0.02$, similar to those of composite AGN/starbursts
in ULIRGs/SMGs at comparable redshifts (Alexander et al.\ 2005). It
therefore appears likely that these galaxies also contain a
dust-enshrouded starburst component, powering at least 80\% of the
far-infrared emission. Correcting for 20\% AGN contribution, we
estimate that the host galaxies have SFRs
$\gtrsim$500$M_\odot$\,yr$^{-1}$ (assuming the far-infrared/SFR
conversion of Kennicutt\ 1998). The host galaxies embedded within
these LABs are probably undergoing an episode of co-eval black-hole
growth and star-formation. Both these processes deposit energy into
the IGM, and for the remainder of this article, we discuss the role of
this heating in powering the extended Ly$\alpha$ emission, and rule
out some other power sources (inverse Compton scattering, cooling)
that have been proposed for LAB formation.

\section{Discussion: what powers LABs?}

Clearly the host galaxies embedded within LABs are extremely
energetic, but can this energy be harnessed to give rise to the
extended Ly$\alpha$ emission?  There are only two basic mechanisms
that transfer the output from the host galaxy into an extended halo:
photo-ionization from UV photons and mechanical feedback. We assess
the viability of each of these power sources in the following
discussion, and conclude with a discussion comparing the physical
viability of cooling versus heating models of LAB formation.

\subsection{Photo-ionization}

When considering photo-ionization, we are only concerned with photons
with $h\nu>13.6$\,eV, and so our constraints on the UV/optical portion
of the SED are important here. The optical/near-infrared photometry
are interpolated using the spectral fitting code {\sc hyperz}
(Bolzonella,\ Miralles \& Pell\'o 2000). Since the IRAC bands are
thought to be contaminated by a hot dust component, we restrict this
fit to $\lambda_{\rm rest} \ls 1$$\mu$m. Figure\ 3 shows the best
fitting SED, which assumed a continuous star formation history of
duration $\sim$100\,Myr (although it is not clear how to interpret
this `age' here; the fit is more useful as an interpolation of the
observed photometry). The UV/optical continuum is the combination of
intrinsic emission from stars and the AGN, attenuated by internal
extinction and (at shorter wavelengths) by foreground Ly$\alpha$
Forest absorption.  However, since some of the self-absorbed radiation
has been re-distributed to other parts of the SED, we can attempt to
reconstruct the intrinsic UV luminosity from massive stars and the AGN
component and assess whether these are sufficient to photo-ionize the
halo.

1)\ {\it AGN contribution---}\ To estimate the intrinsic rest-frame UV
luminosity of the AGN, we apply the simple power-law extrapolation of
Steffen et al.\ (2006): $\alpha_{\rm OX} = 0.3838\log_{10}(\nu l_{\rm
  2keV} / \nu l_{\rm 2500})$. For our typical AGN, extrapolating from
the measured X-ray luminosities, we find $\alpha_{\rm OX}\simeq -1.5$.
We indicate the predicted UV luminosity in Fig.~3. Note that both this
power-law extrapolation and normalized RQQ template of Elvis et al.\
(1994), gives $L_{\rm 2500} \sim 10^{45}$\,erg\,s$^{-1}$. This is an
order of magnitude larger than the observed 2500\AA\ luminosity for
the galaxy, implying strong extinction consistent with the flat X-ray
spectral slopes of the LABs. We assess the role of this obscuration on
the escape of photo-ionizing radiation below.

2)\ {\it Massive stars---}\ The bolometric luminosities of LABs are
dominated by far-infrared emission, and the crude limits on the LABs'
$L_X/L_{\rm FIR}$ suggest that $\sim$20\% of this is likely to be
provided by the AGN (Alexander et al.\ 2005). The remaining power is
predicted to come from dust heated in the UV radiation field of
massive stars, and so to estimate the un-obscured SFR, we convert from
the corrected $L_{\rm FIR}$ (Kennicutt\ 1998). To estimate the
intrinsic UV/optical emission from this starburst component, we scale
the {\it Starburst99} models of Leitherer et al.\ (1999).  The
resulting intrinsic UV--optical SED for a starburst representative of
our composite X-ray detected LAB is shown in Fig.~3 (we assume a Solar
metallicity, Salpeter IMF with upper stellar mass cut-off of
100\,$M_\odot$).  In the absence of obscuration, the intrinsic UV
luminosities are $\sim$2 orders of magnitude larger than the observed
Ly$\alpha$ luminosities, and thus provide an adequate supply of
ionizing photons.

\medskip

In Figure~4 we compare the integrated 200--912\AA~luminosity of the
host galaxies (split into a AGN and starburst component) to the
Ly$\alpha$ luminosity of the LAB. We show that even with small escape
fractions, the luminosity of the AGN/starburst is easily sufficient to
power the LABs' Ly$\alpha$ luminosities via photo-ionization. As Fig~1
shows, there could be quite large variation in the UV escape fraction
from source to source (partly, this could be due to geometric
effects). We attempt to estimate a representative escape fraction
($f_{\rm esc}$) of UV photons from the AGN and starburst components by
comparing the intrinsic UV luminosity for each component to the
observed continuum luminosity at 1500\AA\ (Fig.~3). We make the
assumption that this extinction can also be applied at 912\AA, and
this implies $f_{\rm esc}[{\rm AGN}]\sim0.07$ and $f_{\rm esc}[{\rm
  SF}]\sim0.006$.  On Figure~4 we illustrate the region where
photo-ionization can fully power a LAB taking into account each
$f_{\rm esc}$ -- the reader can scale these lines to test the effect
of various levels of obscuration. All LABs with detected AGN fall in
the region where an AGN, starburst or combination of both can fully
photo-ionize the halo. Similarly, LABs with submm detections
(including the average stacked flux) but no formal X-ray counterpart
are also consistent with an ionizing power source of either starburst
or AGN (Fig~4).

What of the role of extinction on the extended Ly$\alpha$ emission
itself?  Ly$\alpha$ is a resonantly scattered emission line, and so it
is easy to destroy the line in the presence of dust. This is not
likely to be an issue in the LAB halos, since the Ly$\alpha$ photons
are generated in the extended gaseous nebula, well away from the
obscuring material in the host galaxy, although the dust could be
potentially extended on small scales (e.g.\ Matsuda et al.\
2007). Furthermore, radiative transfer could serve to extend the
Ly$\alpha$ emission over larger scales; in fact, in some models where
cooling is invoked to explain the Ly$\alpha$ emission, this is
essential to reproduce LABs on scales of LAB\,1 or LAB\,2 (Fardal et
al.\ 2001).

In addition to direct photo-ionization, the galaxy can inject kinetic
energy in the IGM via outflows. Energy deposited in this way could
also power Ly$\alpha$ emission by promoting collisional
excitation/ionization, or if it is capable of generating a shock, by
photo-ionization.

% FIG 4
\begin{figure}
\centerline{\includegraphics[width=0.5\textwidth]{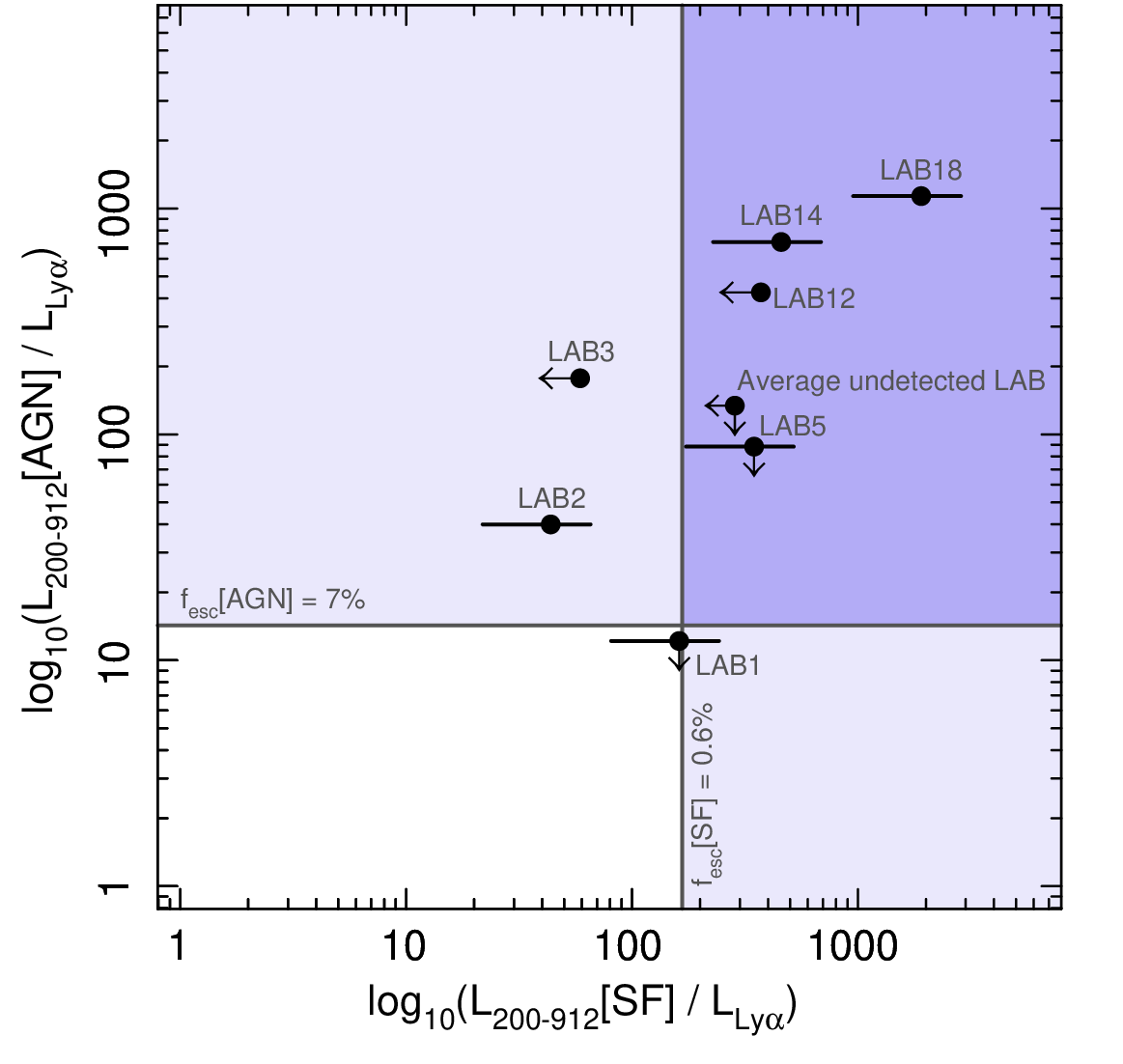}}
\caption{A comparison of the intrinsic UV
  (200--912\AA) luminosities of LABs originating from AGN and
  star-formation. This has been estimated for the AGN and starburst
  components separately (\S4.1) to examine the energetics of each
  component, relative to the total Ly$\alpha$ emission. Although in
  all cases the intrinsic UV luminosities from both components is
  easily sufficient to power the Ly$\alpha$ emission, these have to be
  modified to account for a dust covering fraction, which will
  attenuate the number of ionizing photons. Using the composite SED
  shown in Fig.~3 as a guide, we estimate that the escape fractions
  are $\sim$7\% and $\sim$0.6\% for the AGN and starburst photons
  respectively, and we indicate these fractions on the figure. Note
  that even with this heavy obscuration, photo-ionization is
  sufficient to power the LABs alone.}
\end{figure}

\subsection{Mechanical energy}

As in the photo-ionization models, the total mechanical energy
available to the Ly$\alpha$ halo is derived from both the massive
stars (the detonation of supernovae, and to a lesser extent, stellar
winds) and the AGN (accretion-related outflows). To evaluate the
energy deposited in the IGM by supernovae (stellar winds are not
likely to provide significant feedback, except in very young
starbursts), we again apply the {\it Starburst99} model, but this time
consider mechanical luminosity, rather than UV luminosity. Each
supernova can release $\sim$10$^{51}$\,erg, but only 10\% of this is
believed to pressurize the ISM (Thornton et al.\ 1998) -- the
remainder of the energy is lost to alternative radiative
processes. Assuming the same Salpeter IMF as used in the
photo-ionization calculation, the total mechanical energy from SNe can
be expressed $L_{\rm SNe}/L_{\rm FIR} = 2.7\times10^{-3}$. This simple
scaling assumes that the far-infrared emission is dominated by
star-formation, and that the burst is $\gs$10$^8$ years old. Note that
in our case, we have made a conservative correction for a 20\%
contribution to $L_{\rm FIR}$ from the AGN. We find a large range of
$L_{\rm SNe}/L_{\rm Ly\alpha}$ for the X-ray detected LABs, with
$L_{\rm SNe}/L_{\rm Ly\alpha}\simeq0.4$ for LAB\,2, and $L_{\rm
  SNe}/L_{\rm Ly\alpha}\simeq16$ for LAB\,18. This simply reflects the
fact that smaller, lower luminosity LABs are `easier' to power.

In addition to the output from supernovae, we also have the energy
deposited by outflows from the AGN. A radiation-pressure driven
bi-polar outflow could arise if UV photons deposit momentum in a
covering shell of dust which is then driven out of the
galaxy. Unfortunately, confirming outflows in LABs is extremely
challenging, not only in terms of observational overhead but also
because of the somewhat ambiguous observational signatures of
inflow/outflow (e.g.\ Dijkstra, Haiman \& Spaans 2006). Nevertheless,
the best observational evidence that LABs are experiencing some form
of mechanical feedback is provided by integral field (IFU)
observations of LAB\,1 (Bower et al.\ 2004) and LAB\,2 (Wilman et al.\
2005). LAB\,1 exhibits a chaotic velocity structure and a Ly$\alpha$
`cavity' in the vicinity of the host identified by Geach et al.\
(2007), and LAB\,2 shows evidence of a large scale ($\sim$100\,kpc)
galaxy-wide outflow traced by a Ly$\alpha$ absorption feature with
remarkable velocity coherence. Both of these observations support a
model where mechanical energy is being deposited into the IGM, and
therefore capable of providing power for the extended Ly$\alpha$
emission. 

If one takes both the energy available from photo-ionization and
mechanical deposition (heating), it is clear that the energy supplied
by the LAB host galaxies can be orders of magnitude larger than the
energy released in the Ly$\alpha$ emission. We take this as compelling
evidence that heating must be crucial in powering LABs. In the
remainder of the discussion we examine two other proposed LAB
formation mechanisms: inverse Compton scattering and cooling. We
assess whether these other physical processes are likely to operate in
LABs, compared to the feasibility of the heating model described
above.

\subsection{Extended X-ray emission}

\subsubsection{Inverse Compton scattering}

The IC mechanism -- up-scattering of photons by a population of
relativistic electrons -- becomes more viable as a potential power
source for extended Ly$\alpha$ emission at high redshifts due to the
$(1+z)^4$ evolution in the CMB energy density (e.g.\ Scharf et al.\
2003; Fabian et al.\ 2009).  CMB photons (or far-infrared photons from
the galaxy itself) could be up-scattered to X-ray energies, and then
go on to photo-ionize a halo of neutral hydrogen. Although the LABs
show no current radio activity, this does not rule out a previous
radio-loud mode that could have provided a scattering population of
electrons distributed to several tens of kiloparsec from the
source. Therefore, one way of detecting the IC mechanism at work is to
search for extended X-ray emission.

No individual LABs show evidence for {\it extended} X-ray
emission\footnote{In a previous, shallow (78\,ks) {\it Chandra}
  observation of SS\,A22, Basu-Zych \& Scharf\ (2004) claimed that
  LAB\,2 has some evidence of extended X-ray emission, but we do not
  confirm that result here.}; however, using a summation technique
incorporating the {\it Chandra} data from all 29 LABs, we can search
for an average signal from the extended LAB regions that falls below
the detection threshold of an individual source.  Using the
\hbox{0.5--2~keV} image from Lehmer et al.~(2009), we summed the
source-plus-background counts for pixels within the LAB isophotal
regions as defined by Matsuda et al.~(2004). This stacking technique
provides us with an effective exposure time of $\sim$9.3\,Ms.  In
these summations, we excluded circular regions of radius $2 \times$
the 90\% encircled energy fraction radius for individually detected
point sources. We then extracted and summed background counts from
pixels within the same set of isophotal regions after shifting them by
$\sim$70$''$.  In total, we extracted 84 source-plus-background counts
over 3658 on-source pixels and 83 counts over 4239 off-source pixels.
This gives an on-source fluctuation of $\sim$1.6$\sigma$ above the
background. 

To calculate the X-ray luminosity limit for IC emission, we first
estimated the 3$\sigma$ upper limit on the extracted
source-plus-background counts rescaled to the {\it total} LAB
isophotal area (i.e., the ratio of the total LAB isophotal area and
the area used to extract counts).  Using the vignetting-corrected
0.5--2\,keV exposure map from Lehmer et al.~(2009), we then computed
the total effective exposure for the 29 sources to be $\sim$9.3\,Ms.
This implies a 3$\sigma$ upper limit on the \hbox{0.5--2~keV}
count-rate to be $\ls 4.1 \times10^{-6}$~counts~s$^{-1}$.  Assuming a
power-law spectrum (appropriate for X-ray emission from IC scattering;
$\Gamma=1.7$, see Scharf et al.\ 2003), we find a 2--8\,keV luminosity
limit of $L_X<$$1.5\times10^{42}$\,erg\,s$^{-1}$ (3$\sigma$).  We
conclude that the lack of extended X-ray emission around the LABs
rules out the IC mechanism as a viable power source.

\subsubsection{Hot gas component}

In the classic picture of galaxy formation, gas entering dark matter
halos can be shock heated to the virial temperature of the halo (White
\& Frenk\ 1991). Our non-detection of extended X-ray emission around
LABs provides a useful limit on the thermal properties of the gas
halo, and therefore we are able to speculate about the properties of
the dark matter halos that LABs inhabit. 

For example, the virial temperature of a halo of mass $10^{13}M_\odot$
is $T \sim 10^7$\,K.  Using the upper limit on the X-ray count-rate
described above and assuming a Raymond-Smith plasma SED with $Z = 0.2$
implies a rest-frame \hbox{0.5--2~keV} luminosity limit of $<$$2
\times 10^{43}$~erg~s$^{-1}$.  If gas was cooling from the virial
temperature, then we would expect $L_X/L_{\rm Ly\alpha} \gs 10^3$
(Cowie, Fabian \& Nulsen [1980]; Bower et al. [2004]). We find
$L_X/L_{\rm Ly\alpha} \ls 1$, and so our observations imply there is
no hot ($10^7$\,K) gas component in these halos.

This measurement does not rule out a `cold' cooling mode in the galaxy
halo (see Fardal et al.\ 2001). Can such cooling radiation be a viable
power source for the LABs? In the final discussion we investigate the
likelihood for this scenario, compared to the picture where LABs are
powered by heating by the embedded host galaxy.

\subsection{Cooling versus heating: which wins?}

The simple cooling of gas within dark matter halos has been used to
explain the existence of LABs not containing obvious `active' galaxies
such as those presented here (e.g.\ Smith\ \& Jarvis\ 2007). The best
candidate for a LAB powered by cooling was identified by Nilsson\ et\
al.\ (2006), in the GOODS-South field. While Nilsson\ et\ al.\ do not
associate the GOODS-S LAB ($z=3.16$) with a companion continuum
source, we note there is an IRAC 8$\mu$m, and MIPS 24$\mu$m-detected
source just 3$''$ (20\,kpc) away from the GOODS-S LAB, with a
photometric redshift consistent with the LAB itself. As with some of
the LAB counterparts in this work, this could be a starburst galaxy
offset from the peak of the Ly$\alpha$ emission.  Nevertheless,
Nilsson et al.\ argue that this object is not associated with the LAB,
and in the absence of a detectable ionizing source within the halo
they conclude that cold accretion is the most plausible power
source. What are the physical consequences that must be considered if
cooling flows power LABs? The major hurdle that cooling models must
overcome is the fact that the expected cooling times of these halos is
very short, and this has some profound physical implications regarding
the evolution of the host galaxy. We will illustrate this using a
simple model.

Consider a LAB modelled as an isothermal sphere of gas. The cooling
timescale of this gas halo is simply the ratio of the thermal energy
to the cooling rate, $\Lambda$: $t_{\rm cool} = 3NkT/2n_en_i
\Lambda$. Let us model a primordial gas mixture in collisional
equilibrium as a conservative case (we ignore all other sources of
photo-ionization and cooling via metal lines).  If we assume that {\it
  all} of the cooling is emerging in the Ly$\alpha$ line at the peak
of the cooling function (i.e. $T\sim2\times10^4$\,K; Katz, Weinberg \&
Hernquist\ 1996), then we can estimate the total thermal energy and
therefore cooling timescale of the LAB. This is probably a reasonable
assumption, because as we have seen, there is observational evidence
that suggests gas in the IGM is not in a hot mode (\S4.3.2). This is
also in agreement with theoretical models which suggest that gas
falling into dark matter halos never reaches the virial temperature,
and instead is dominated by a cold mode of accretion, with gas at
$T\sim10^4$\,K (Fardal et al.\ 2001; Haiman \& Rees\ 2001; Kay et al.\
2000; Birnboim \& Dekel\ 2003).  Taking LAB\,2 as a representative
example, the gas halo will lose all of its thermal energy (and
therefore vanish) within $\sim$1.5\,Myr. In order to sustain the LAB
in this cooling model, it follows that one must replenish the warm gas
in the halo as it is being cooled onto the host galaxy. Is this
realistic?

The total mass of material required to pass through this cooling phase
can be estimated by comparing the cooling rate with the likely
lifetimes of LABs. Unfortunately we have no constraints on LABs'
lifetimes, so we make some estimates based on a simple evolutionary
and duty-cycle argument. We know that LABs are commonly associated
with LBGs, and it is not an unreasonable assumption that all LBGs go
through a LAB phase. In SSA\,22a (the LBG survey region of Steidel et
al.\ 2003), $3.5\pm1.5$\% of LBGs are associated with LABs. LABs have
been detected over $2.3\ls z\ls6.7$, a span of $\sim$2\,Gyr in cosmic
time (e.g. Smith \& Jarvis\ 2007, Ouchi et al.\ 2008; Yang et al.\
2008), a simple duty cycle argument then implies that the LAB lifetime
is $\sim$50--100\,Myr. Hence if LAB\,2 was to be completely powered by
cooling, then over this duration the central galaxy would have to
accrete $\sim$$10^{12}M_\odot$ of molecular gas.

Bearing in mind that the stellar masses of the LAB hosts are already
$\sim$$10^{11}M_\odot$ (Geach et al.\ 2007; Smith et al.\ 2008;
Uchimoto et al.\ 2008), it seems unlikely that they would increase
their stellar mass by a factor 10$\times$ in such a short period of
time without triggering starburst or AGN activity that would
potentially heat their halos. Nevertheless, some current LAB formation
models propose that the host can be ineffective at influencing the
cold flow in any way. For example, recent high-resolution hydrodynamic
simulations of cold mode cooling in $\sim$$10^{12-13}M_\odot$ halos
suggest that cold ($10^4$\,K) gas enters the galaxy in thin filaments
(Dijkstra \& Loeb\ 2009). The key difference between filamentary cold
flows and the simple isotropic cooling we discussed above is that the
gas enters the galaxy in dense (1--100\,cm$^{-3}$) streams with a
small volume filling factor. The high H{\sc i} densities will shield
the majority of the gas from external ionizing radiation (i.e.\ the
AGN/starburst), and the small angular covering factor means that
terminating the flow via feedback is ineffective, since outflows
emerge from the galaxy through low density patches between the
streams. However, it should be noted that the physical interaction
between AGN/starburst feedback and filamentary cold flows is still
unclear.

As in our simple case, the main problem that this refined cooling
model faces is the requirement that a large mass of gas must be
accreted onto a $M_\star \sim 10^{11}M_\odot$ galaxy (the filamentary
cooling mode has a duty cycle of unity, Dijkstra \& Loeb\
2009). Cessation of the cold flow occurs when the halo reaches a
critical mass, which is a function of redshift such that cold flows
terminate by $z<2$ (Dekel et al.\ 2009). Still, at $z\sim3$, this
`over cooling' is exactly the scenario that modern models of galaxy
formation attempt to prevent -- run-away star formation resulting in
too many very massive galaxies. Without introducing feedback that can
terminate cooling, models severely over-predict the number of massive
galaxies at $z=0$ (Bower et al.\ 2006).

\section{Summary \& Final Remarks}

In this deep {\it Chandra} survey of 29 LABs in the SSA\,22
protocluster at $z=3.09$, we have unambiguously identified 5
moderately luminous ($L_{\rm 2-32keV}\sim10^{44}$\,erg\,s$^{-1}$) AGN
embedded within LABs. The high AGN fraction, $17^{+12}_{-7}$\% hints
that an active host galaxy is important for LAB formation, and our
analysis concentrates on how the energetics of the host galaxies could
relate to the extended Ly$\alpha$ emission. Our main results and
conclusions are:

\smallskip

1)~All five AGN have hard spectral indices, implying intrinsic
obscuring column densities of $N_{\rm H}\gs$10$^{23}$\,cm$^{-3}$.  and
all of the X-ray detected LABs have 8$\mu$m counterparts, implying
rest-frame near-infrared colours consistent with a power-law continuum
associated with warm dust emission (Webb et al.\ 2008). These X-ray
un-detected LABs also have AGN-like near-infrared colours hinting that
they also contain buried AGN (Geach et al.\ 2007; Webb et al.\
2009). Our derived AGN fraction should be considered a lower limit,
and could be as high as $\sim$30\% (or greater) if the AGN are heavily
obscured, or there are a larger population of lower-luminosity AGN.

2)~The intrinsic UV luminosity of the host galaxies (arising from
massive stars and the AGN) is easily sufficient to power the LABs via
photo-ionization, even with large dust covering fractions. When one
includes energy deposited by mechanical feedback it is clear that the
host galaxies can provide all the energy required to explain the
extended Ly$\alpha$ luminosity of LABs.

3)~We find no evidence of extended X-ray emission around the LABs,
ruling out inverse Compton scattering as an important power source for
LABs. Our derived limit on the diffuse X-ray component compared to
extended Ly$\alpha$ luminosity, $L_X/L_{\rm Ly\alpha}\ls 1$, also
implies that there is little or no shock-heated gas at temperatures of
$\sim$10$^7$\,K in the LABs. This crude temperature limit hints that
LABs probably occupy dark matter halos of mass $\ls$$10^{13}M_\odot$.

\smallskip

\noindent Our results strongly support the heating model of LABs,
where the active host is powering the extended Ly$\alpha$ emission,
rather than the so-called `cold accretion' models of LAB
formation. The exact evolutionary history of LABs remains unclear;
however LABs' association with luminous host galaxies is a compelling
hint that they are linked to feedback events at the sites of formation
of massive galaxies and AGN. Admittedly not all LABs show unambiguous
signs of intense starburst or AGN activity, but we feel that this
should not be taken as evidence that cold accretion is at play: the
potentially luminous embedded sources are likely to be heavily
obscured (Geach et al.\ 2007; Webb et al.\ 2009), or fall just below
the sensitivity of current instrumentation (Geach et al.\
2005). Although cooling must occur at some point in LABs' history, any
vestigial cooling must now be overwhelmed by feedback from the galaxy
itself.

In summary, there is little compelling observational evidence
supporting the cooling model. We have shown that in order to power a
LAB by cold accretion over a reasonable time-scale, then the final
mass of the galaxy becomes unreasonably large. This is exactly the
problem that contemporary models of galaxy formation have to overcome:
cooling must be swiftly curbed to prevent a `run-away' star formation
episode resulting in too many massive ($>$$L_\star$) galaxies by $z=0$
(Bower et al.\ 2006). It is possible that LABs could be the epitome of
this physical model of galaxy evolution.

\section*{Acknowledgements}

We thank the referee for helpful comments, and we appreciate useful
discussions with Mark Dijkstra, Chris Done, Caryl Gronwall, Cedric
Lacey and Tom Theuns. J.E.G. is funded by the U.K. Science and
Technology Facilities Council (S.T.F.C.). D.M.A. acknowledges the
Royal Society and the Leverhulme Trust for financial
support. B.D.L. is supported by a S.T.F.C. post-doctoral
fellowship. I.S. acknowledges support from the Royal Society and
S.T.F.C. Additional support for this work was provided by NASA through
{\it Chandra} Award Number SAO G07-8138C (S.C.C., C.A.S., M.V.) issued
by the {\it Chandra} X-ray Observatory Center, which is operated by
the Smithsonian Astrophysical Observatory under a NASA contract.

\end{document}